\documentclass{amsart}

\usepackage{epsfig}

\newcommand{\be}{\begin{eqnarray}}
\newcommand{\ee}{\end{eqnarray}}

\def\ra{\rangle}

\begin{document}

\title{An Explicit Universal Gate-set  for Exchange-Only Quantum Computation}

\author{M. Hsieh$^{1}$, J. Kempe$^{1,2,3}$, S. Myrgren$^1$ and K. B. Whaley$^1$}

\address{$^1$ Department of Chemistry, $^2$ Computer Science Division, University of California, Berkeley\\$^3$ CNRS-LRI, UMR 8623, Universit\'e de Paris-Sud, 91405 Orsay, France }

\date{\today}

\begin{abstract}
  A single physical interaction might not be universal for quantum
computation in general. It has been shown, however, that in some cases
it can achieve universal quantum computation over a subspace. For
example, by encoding logical qubits into arrays of multiple physical
qubits, a single isotropic or anisotropic exchange interaction can
generate a universal logical gate-set. Recently, encoded universality
for the exchange interaction was explicitly demonstrated on
three-qubit arrays, the smallest nontrivial encoding. We now present
the exact specification of a discrete universal logical gate-set on
four-qubit arrays. We show how to implement the single qubit
operations exactly with at most $3$ nearest neighbor exchange
operations and how to generate the encoded controlled-NOT with $27$
parallel nearest neighbor exchange interactions or $50$ serial gates,
obtained from extensive numerical optimization using genetic
algorithms and Nelder-Mead searches. We also give gate-switching times for the three-qubit encoding to much higher accuracy than previously and provide the full specification for exact $CNOT$ for this encoding. Our gate-sequences are
immediately applicable to implementations of quantum circuits with the
exchange interaction.

\end{abstract}
\maketitle

\section{Introduction}

To implement universal computation in the quantum regime, one must be
able to generate any unitary transformation on the logical qubit
states. By now it has become part of the quantum computation folklore
that the group $SU(2)$ of single-qubit operations and an entangling
two-qubit operation such as the controlled-NOT ($CNOT$) can generate
any unitary transformation exactly
\cite{DiVincenzo:95a,Barenco:95a}. Furthermore it has been shown that
there are {\em discrete} universal elementary gate-sets which
approximate any unitary transformation with arbitrary precision
efficiently\footnote{We use {\em efficient} in the computational
sense, meaning that we can implement the transformation with a number
of elementary gates polynomial in the number of qubits. Note that not
all general unitary transformations can be implemented efficiently; in
fact the generic unitary transformation on $n$ qubits requires an
exponential amount of elementary gates. Our usage of {\em efficient}
here means that given there is a sequence of one- and two-qubit gates
that generates $U$ then we can approximate this $U$ to arbitrary
accuracy with a sequence of gates drawn from our elementary set and
such that we only have polynomial overhead in the number of gates
used. Further, to double the precision we only need a constant amount
of additional gates. This is the notion we need to define efficient
computation.} (see \cite{Kitaev:book,Nielsen:book} for details). One
such set is comprised of $\{H,\frac{\pi}{8},CNOT\}$ \cite{Boykin:99a},
where $H$ is the Hadamard transform and $\frac{\pi}{8}$ is a phase
gate, both acting on a single qubit.  In this sense, $H$,
$\frac{\pi}{8}$ and $CNOT$ comprise a quantum analogue to a classical
\textit{universal logical gate-set}.

The traditional paradigm of quantum computation of ``one physical
qubit = one logical qubit'' is often hard to implement because in the
presently known menu of physical implementation schemes, it is usually
difficult to control at least one of either the single-body or the
two-body operations \cite{Kempe:02a}.  

A prime example is the Heisenberg interaction (with Hamiltonian
$H_E^{i,j}=J_{ij} \vec{S}_i \otimes \vec{S}_j$ between spin particles
$i$ and $j$, where $\vec{S}_i=\frac{1}{2} \vec{\sigma}^i$ and $\sigma^i_{x,y,z}$ are the usual Pauli matrices acting on qubit $i$). It has many attractive features
\cite{Loss:98a,Burkard:99b} that have led to its being chosen as the
fundamental two-qubit interaction in a large number of recent
proposals: Its functional form is very accurate --- deviations from
the isotropic form of the interaction, arising only from relativistic
corrections, can be very small in suitably chosen systems.  It is a
strong interaction, so that it should permit very fast gate operation,
well into the GHz range for several of the proposals. At the same
time, it is very short ranged, arising from the spatial overlap of
electronic wavefunctions, so that it should be possible to have an
on-off ratio of many orders of magnitude. We will assume that the
interaction can be switched on and off between coupled qubits
\cite{Burkard:99b}. Unfortunately, the Heisenberg interaction by
itself is not a universal gate, in the sense that it cannot generate
any arbitrary unitary transformation on a collection of spin-1/2
qubits.  So, every proposal has supplemented the Heisenberg
interaction with some other means of applying independent one-qubit
gates (which can be thought of as time-dependent local magnetic
fields).  But the need to add this capability to the device adds
considerably to the complexity of the structures, by putting
unprecedented demands on ``g-factor'' engineering of heterostructure
materials \cite{DiVincenzo:99a,Vrijen:00a}, requiring that strong,
inhomogeneous magnetic fields be applied, or involving microwave
manipulations of the spins that may be slow and may cause heating of
the device. These added complexities may well exact a high cost,
perhaps degrading the quantum coherence and clock rate of these
devices by several orders of magnitude.

\textit{Encoded universality} \cite{Kempe:01b} provides a way around
this problem in some crucial cases, for example when the ``easy''
interaction is the exchange interaction, by entirely eliminating the
need for single-body physical operations. By encoding each logical
qubit in an array of multiple physical qubits, sequences of two-body
nearest-neighbor exchange interactions are sufficient to generate the
logical $SU(2)$ and $CNOT$ operations\footnote{Note that it has been
shown that a {\em generic} two-qubit interaction alone generates
universal computation \cite{Deutsch:95a,Lloyd:95a}. However, by an irony
of nature most implementable interactions in current quantum
computation schemes happen to fall in the set of exceptions to
this. These exceptions include the ubiquitous exchange interaction
(both isotropic and anisotropic) and several other interactions that
exhibit a certain amount of symmetry, which makes them {\em
non-generic} in the above sense. Even for interactions that fall into
the category of being universal by themselves, explicit
gate-constructions have to be found in a case by case basis.} on the
encoded qubits \cite{Kempe:01a,Kempe:01b,Bacon:99b} and single-spin
operations and all their attendant difficulties can be avoided.

One drawback of the theory of encoded universality \cite{Kempe:01a} is
that it establishes the sufficiency of certain two-body interactions
for universality in a non-constructive way, not offering explicit
methods with which to specify the sequences of physical implementable
Hamiltonians corresponding to the encoded logical gates. In particular
it is not clear at the outset how many physical interactions are
required to implement each of the logical gates in some layout of the
qubits. Encoded computation schemes are only viable if the number of
physical interactions to be applied to the qubits is not too large,
where the threshold is determined by currently achievable decoherence
and switching times. In most cases, numerical methods are the only way
to find explicit sequences of Hamiltonians for a set of universal
gates for some realistic arrangement of the physical qubits. Recently,
more or less explicit universal logical gate-sets have been given for
a three-qubit encoding using only the exchange interaction
\cite{DiVincenzo:00a}, for the $XY$-interaction \cite{Kempe:02a} and
for the generalized anisotropic exchange Hamiltonian \cite{Vala:02a}.
In \cite{DiVincenzo:00a} an initial encoding of three physical qubits
per logical qubit is used and a sequence of 19 Hamiltonians is
presented that implements the encoded $CNOT$. However this $CNOT$ is given
up to local unitary operations only, and the encoded single-qubit
operations are given in terms of Euler-angle rotations for the group
$SU(2)$. Some further processing is needed to obtain a universal
discrete gate-set, needed to implement quantum circuits {\it in terms of the computational basis} \cite{Myrgren:03a}.

We present here a complete scheme for universal quantum computation on
four-qubit encodings in a one- (or two-)dimensional layout with
nearest neighbor interactions only.  We specify the encoding and
layout and give all the gate switching times to obtain the encoded
$H$, $\frac{\pi}{8}$ and $CNOT$ in the computational basis without
further post-processing. This scheme provides an immediately
applicable building block for exchange-only quantum circuits.  We also
provide new gate sequences for the $CNOT$ in the three-qubit encoding
to higher precision and with different symmetries than in
\cite{DiVincenzo:00a} and provide the complete set of gates for the
exact encoded $CNOT$ in this smaller encoding.  Although the
four-qubit encoding has a slightly larger overhead in spatial
resources than the three-qubit encoding it offers several
advantages. A quantum computation begins by setting all encoded qubits
to the (logical) zero state. In our scheme this state is a tensor
product of singlet states.  This state is easily obtained using the
exchange interaction: if a strong $H_{12}$ is turned on in each coded
block and the temperature made lower than the strength $J$ of the
interaction, these two spins will equilibrate to their ground state,
which is the singlet state. Unlike the smaller three-qubit encoding we
do not require here any additional weak magnetic fields for
initialization.  This aspect renders the four-qubit encoding
particularly attractive. Another advantage is that the four-qubit
scheme is conceptually simpler for use in quantum logic when the
properties of robustness to noise are also taken into account.
Whereas the four-qubit logical states constitute a decoherence free
subspace (DFS) under collective decoherence \cite{Kempe:01a}, the
three-qubit logical states constitute a decoherence free subsystem in
which the logical state evolution is defined by only one component of
the tensor product space.  A third advantage of the four-qubit
encoding is that additional protection against single qubit errors can
be achieved in this case by control of extra exchange interactions to
form a supercoherent qubit \cite{Bacon:01b}.

The structure of this paper is as follows.  First, we describe the
four-qubit encodings which define our logical space and give the
physical layout of the qubits. We then specify the Hamiltonians
required to generate the single-qubit operations and the $CNOT$ gate
on the encoded qubits.  The numerical methods used to obtain the
exchange gate sequence for the encoded $CNOT$ and further details are
described in Appendices \ref{App:1} and \ref{App:2}. The same numerical methods
are used to obtain new high accuracy gate sequences for the exact
$CNOT$ in the three-qubit encoding.  We conclude with a brief
discussion of a solid-state implementation scheme in which these
results can be readily applied, and some open problems meriting
further consideration.

\section{Four-qubit encoding}

We define the logical zero-state and one-state on one encoded
qubit as
\begin{eqnarray}
|0_{L}\rangle = \frac{1}{2}\left(|01\rangle - |10\rangle
\right)\otimes\left(|01\rangle - |10\rangle \right)
\label{E:encoding1} \\
|1_{L}\rangle =
\frac{1}{\sqrt{3}}\left(|t_{+}\rangle\otimes|t_{-}\rangle -
|t_{0}\rangle\otimes|t_{0}\rangle
+|t_{-}\rangle\otimes|t_{+}\rangle \right), \label{E:encoding2}
\end{eqnarray}
where $|t_{0}\rangle = \frac{1}{\sqrt{2}}\left(|01\rangle +
|10\rangle\right)$, $|t_{-}\rangle = |00\rangle$, and $|t_{+}\rangle =
|11\rangle$.  For a more detailed discussion on how to obtain  these
encodings, refer to \cite{Kempe:01b,Bacon:99b}.  To initialize a
computation all logical qubits have to prepared in the $|0_L\ra$
state. Note that here the $|0_L\ra$ state is a tensor product of
singlets $\frac{1}{\sqrt{2}} (|01\ra - |10\ra)$. As detailed in the
previous section this will be advantageous in many experimental
settings since it will permit easy initializion of the logical qubits
at the beginning of a computation.

The arrays are spatially configured to permit serial nearest-neighbor
exchanges between the physical qubits in a one-dimensional layout:

\begin{figure}[h]\label{QubitLayout}
\begin{picture}(0,80)
\includegraphics{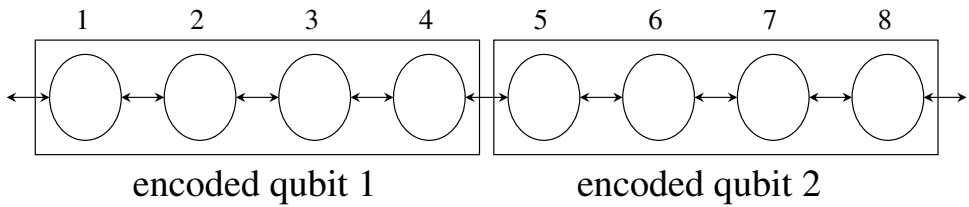}
\end{picture}
\caption{Serial configuration of eight physical qubits comprising
a system of two encoded qubits.}
\end{figure}

We could also imagine these qubit arrays in a two-dimensional layout,
where several of the one-dimensional layers are stacked on top of each
other. Our construction of gate sequences will also hold for the
two-dimensional setting. It suffices to note that the only difference
is that along the layers of arrays the fourth qubit of each array is
coupled to the first qubit of the next, whereas between layers the
first qubit of one array in one layer is interacting with the first
qubit of an array in the other. But note that both encoded states
$|0_L\ra,|1_L\ra$ are symmetric with respect to swapping qubit $1$
with qubit $4$ and qubit $2$ with qubit $3$, so we obtain exactly the
same gate-sequence for $CNOT$ for a coupling of the two first qubits of
an array - all we need to do is to relabel the qubits on the bottom
array as $4,3,2,1$.

The basis states of the logical space defined by two encoded qubits
are $|0_{L} 0_{L}\rangle$, $|0_{L} 1_{L}\rangle$, $|1_{L}
0_{L}\rangle$ and $|1_{L} 1_{L}\rangle$. 

\section{Single-Qubit Operations}

Our goal is to construct the single-qubit Hadamard $H$ and $\frac{\pi}{8}$ 
 gates, defined as
\begin{equation}
\frac{\pi}{8} = e^{i\pi/8} \left( \begin{array}{cc} e^{-i\pi/8} & 0 \\
0 & e^{i\pi/8}  \end{array} \right),~H = \frac{1}{\sqrt{2}} \left( \begin{array}{cc} 1 & 1 \\
1 & -1  \end{array} \right). \label{E:HadPi8}
\end{equation}
on the encoded qubits, using a sequence of exchange interactions
$H_E^{i,i+1}=\frac{J_{i,i+1}}{4} (\sigma_{x}^i\sigma_{x}^{i+1} + \sigma_{y}^i\sigma_{y}^{i+1} + \sigma_{z}^i\sigma_{z}^{i+1})$ on
adjacent pairs of physical qubits $i$ and $i+1$. The matrices
$\sigma_{x,y,z}^i$ are the usual Pauli matrices acting on the $i$th
qubit and $J_{i,i+1}$ is the coupling constant. When we write $H_E^{i,i+1}$ we assume that the Hamiltonian acts
on the $i$th and $i+1$st qubit as specified and as the identity on all
the other qubits. For convenience we are going to add a multiple of the identity to $H_E$ (which just gives an unobservable global phase), and work
with the rescaled interaction
\begin{equation}
E^{i,i+1}=\frac{1}{2}(\sigma_x^i \otimes \sigma_x^{i+1} +\sigma_y^i \otimes \sigma_y^{i+1} +\sigma_z^i \otimes \sigma_z^{i+1} +I^i \otimes I^{i+1})=
\left(
\begin{array}{llll} 1 & 0 & 0 & 0 \\ 0 & 0 & 1 & 0\\0 & 1 & 0 & 0\\ 0 & 0 & 0 & 1
\end{array} 
\right). \label{E:Hexchange}
\end{equation}
so that $exp \left( - (J_{ij}/\hbar)t {\bar S}_i \cdot {\bar S}_j \right) \equiv exp \left( -i(J_{ij}/2\hbar)t E^{i,j} \right)$ (up to a global phase).
We will give all exchange times in units of $2\hbar/J_{ij}$  \cite{Loss:98a}.

Consider the effect of two particular exchanges on the logical states
of a single encoded qubit. First, we note that in the code-subspace
the action of the exchange $E^{1,2}$ is equal to that of $E^{3,4}$, with both of these generating the transformation
$|0_{L}\rangle \rightarrow -|0_{L}\rangle$ and $|1_{L}\rangle
\rightarrow |1_{L}\rangle$.  So -$E^{1,2}$ is equivalent to a
$\sigma_{z}$ operation on a single logical qubit. Therefore the
Hamiltonian for the encoded $\frac{\pi}{8}$ operation, up to an
unobservable global phase, is exactly $e^{i\frac{\pi}{8}E^{1,2}}$.

Next, consider the exchange $E^{2,3}$. The action of this in the code-space is equivalent to $E^{1,4}$.
The effect of the $E_{23}$ operation on a single logical qubit is:
\begin{equation}
E^{2,3} =  \left(
\begin{array}{cc} \frac{1}{2} & \frac{\sqrt{3}}{2}\\ \frac{\sqrt{3}}{2} &
-\frac{1}{2}\\ \end{array} \right). \label{E:SigXcomp}
\end{equation}

By examining the effect of the $E^{1,2}$ and $E^{2,3}$ operations on
the code-space, we can obtain an exact encoded Hadamard operation as $H
= e^{it_{1}E^{1,2}}e^{it_{2}E^{2,3}}e^{it_{1}E^{1,2}}$, where
$t_{1} = \frac{1}{2}\arcsin\sqrt{\frac{2}{3}}=0.4777$ and $t_{2}
= \arccos{\sqrt{\frac{1}{3}}}=0.9553$.  

The exact specifications of exchange operations for
these single qubit gates are depicted in
Fig. 2.

\begin{figure}[h]\label{OneQubitOperations}
\begin{picture}(0,130)
\includegraphics{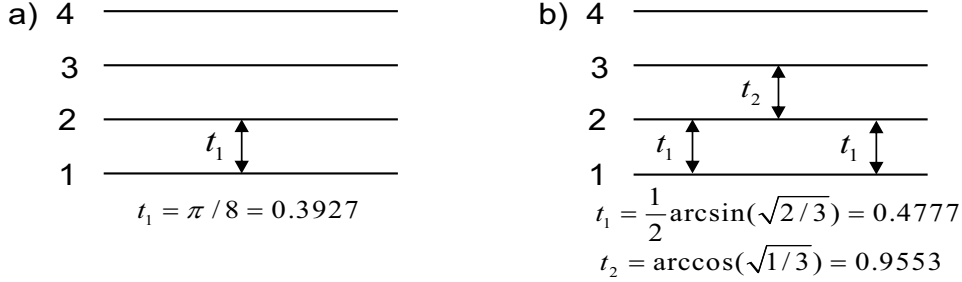}
\end{picture}
\caption{a) One exchange interaction is sufficient to generate the
encoded $\frac{\pi}{8}$ gate.  b) three 
nearest neighbor exchanges allow to generate the encoded Hadamard gate. The
$t$ values are the time parameters corresponding to the individual
exchanges with $t=\pi/8=0.3927$, $t_1=0.4777$ and
$t_2=0.9553$. 
All times are given in units of $2 \hbar/ J_{ij}$.}
\end{figure}

\section{Encoded $CNOT$ operation}\label{sec:CNOT}

To obtain the encoded $CNOT$ we used numerical methods and proceeded in
two stages. In the first step we attempted to obtain a gate $U_{cnot}^{exchange}$ that is
{\em equivalent} to the encoded $CNOT$ {\em up to local unitary
transformations} on the encoded qubits. $U_{cnot}^{exchange}$ is {\em locally
equivalent} to $CNOT$ if there are single-qubit unitary operations
$U_1,U_2,V_1,V_2$ (acting on the first and second encoded qubit,
respectively) such that
\begin{equation}\label{locequivalent}
CNOT = (U_1 \otimes U_2) U_{cnot}^{exchange} (V_1 \otimes V_2)
\end{equation}
In the second stage we numerically obtained the local unitary operations
$U_1,U_2,V_1,V_2$ to get from $U_{cnot}^{exchange}$ to the real $CNOT$ in the computational 
basis and to obtain the gate-sequences
of exchange interactions corresponding to each of these local operations $U_i,V_i$, $i=1,2$. The reason for
splitting the task into these two stages is the following.  A result by
Makhlin \cite{Makhlin:00a} shows that all locally equivalent gates are
characterized by only three real parameters, $M_1$ (a complex number) and $M_2$ (real), which we
will refer to as the {\em Makhlin-invariants} in what follows. Appendix
\ref{App:1} gives a brief summary of how $M_1$ and $M_2$ are
calculated. For the $CNOT$, $M_1=0$ and $M_2=1$. The reduction to three parameters greatly simplifies the
numerical search and allowed us to obtain the gate sequences by a
combination of genetic algorithms \cite{Buckles:book} and Nelder-Mead simplex
searches \cite{Lagarias:95a,Nelder:65a}. The Makhlin invariants give rise to a simple fitness
function $f = \|M_{1}(CNOT) - M_{1}(Candidate)\| + \|M_{2}(CNOT) -
M_{2}(Candidate)\|$, where $\| .\|$ is the complex norm.  A gate sequence that generates a value $f=0$ is
therefore equivalent to a $CNOT$.

It is imperative for succesful quantum
computation over a subspace that any permissible operation over the
encoded qubits must act unitarily on linear combinations of these
basis states and not ``leak'' any amplitude out of the subspace into
its complement and vice versa. We will capture this requirement by defining a
\textit{leakage parameter} $\Lambda$. Then any permissible two-qubit
physical operation $W$ must keep the code-subspace invariant,
i.e. obey the following equation\footnote{Note that exchange
operations {\em within} an array of $4$ qubits keep the one-qubit code
space invariant and do not leak. Leakage can only occur when we couple
two arrays (see \cite{Kempe:01a} for details).}:
\begin{eqnarray}
 \Lambda = 4-\sum_{s=0}^{1}\sum_{t=0}^{1}\sum_{u=0}^{1}\sum_{v=0}^{1} |\langle
s_{L} t_{L}| W |u_{L} v_{L} \rangle|^{2} = 0. \label{E:Leakage}
\end{eqnarray}
If leakage occurs $\Lambda >0$. 
We note that any $W$ that is locally equivalent to $CNOT$ or any
other unitary logical operation over two encoded qubits must
\textit{by definition} generate a leakage parameter of $\Lambda=0$. 
However, we have found through numerical inspection that the
search space in this problem is heavily pocked with local minima
in which the Makhlin invariants are close to the desired values for $CNOT$,
but which ``leak'' out of the logical space by generating a
$\Lambda$ value larger than $0$. In our numerical searches, we
overcame this problem by defining our fitness function as
$\mathcal{F}$ = f + $\Lambda$, optimizing explicitly for not
only a Makhlin invariant match, but also for non-leakage.

 A detailed description of our search algorithms and of the accuracy
 of our gate sequence can be found in Appendix \ref{App:2}.

 For a gate $U_{cnot}^{exchange}$, locally equivalent to the $CNOT$, we found a gate sequence of
 $34$ nearest neighbor exchange interactions. Figure 3 and Table \ref{tab:4qbcnot}
 show the layout and time parameters.
\begin{figure}[htbp]\label{CNOTcircuit}
\begin{picture}(0,170)
\includegraphics{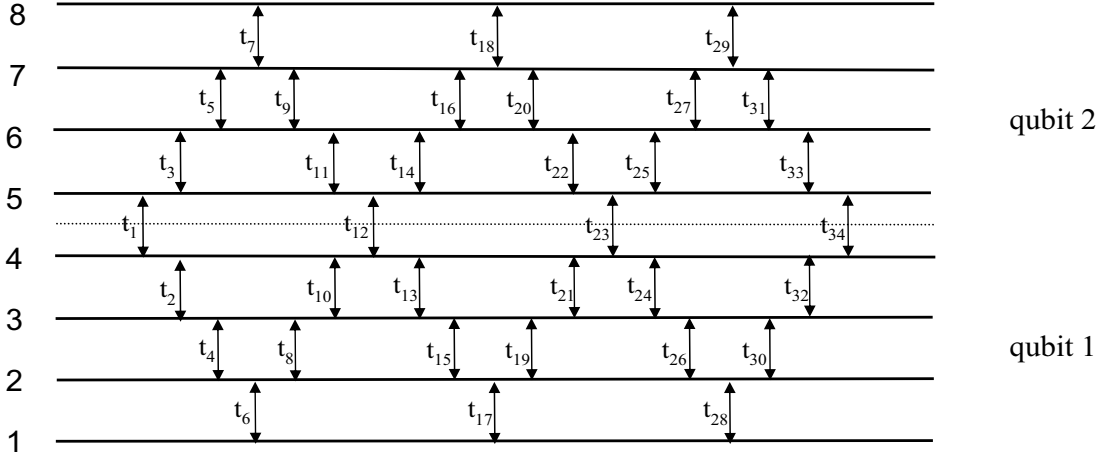}
\end{picture}
\caption{Gate sequence of $34$ exchange interactions for the encoded $CNOT$, with
the corresponding time parameters, given in Table \ref{tab:4qbcnot}. }
\end{figure}

\begin{table}[ht]
\begin{tabular}{|c|c|c|c|c|c|c|c|}\hline 
Exchange & Qubit & Qubit &          & Exchange   &  Qubit & Qubit &   \\
Time     &    1    &   2    & Times &   Time     &    1   &   2   & Times \\
\hline \hline
$t_{1}$  &    4    &   5    & 1.90680 &  $t_{18}$ &  7    &   8   & 0.95629 \\
$t_{2}$  &    3    &   4    & 1.59536 &  $t_{19}$ &  2    &   3   & 1.06260 \\
$t_{3}$  &    5    &   6    & 1.26290 &  $t_{20}$ &  6    &   7   & 0.68131 \\
$t_{4}$  &    2    &   3    & 1.59745 &  $t_{21}$ &  3    &   4   & 0.59800 \\
$t_{5}$  &    6    &   7    & 2.06920 &  $t_{22}$ &  5    &   6   & 1.19942 \\
$t_{6}$  &    1    &   2    & 0.05331 &  $t_{23}$ &  4    &   5   & 1.04719 \\
$t_{7}$  &    7    &   8    & 0.76951 &  $t_{24}$ &  3    &   4   & 3.14138 \\
$t_{8}$  &    2    &   3    & 1.59747 &  $t_{25}$ &  5    &   6   & 0.95529 \\
$t_{9}$  &    6    &   7    & 0.71337 &  $t_{26}$ &  2    &   3   & 1.63957 \\
$t_{10}$ &    3    &   4    & 1.59958 &  $t_{27}$ &  6    &   7   & 1.91303 \\
$t_{11}$ &    5    &   6    & 1.26287 &  $t_{28}$ &  1    &   2   & 2.47920 \\
$t_{12}$ &    4    &   5    & 1.90667 &  $t_{29}$ &  7    &   8   & 2.18627 \\
$t_{13}$ &    3    &   4    & 0.59810 &  $t_{30}$ &  2    &   3   & 1.05736 \\
$t_{14}$ &    5    &   6    & 1.71467 &  $t_{31}$ &  6    &   7   & 0.94814 \\
$t_{15}$ &    2    &   3    & 1.06264 &  $t_{32}$ &  3    &   4   & 3.14170 \\
$t_{16}$ &    6    &   7    & 0.91559 &  $t_{33}$ &  5    &   6   & 4.09690 \\
$t_{17}$ &    1    &   2    & 2.30240 &  $t_{34}$ &  4    &   5   & 2.09434 \\
\hline
\end{tabular}
\vskip.1cm
\caption{Gate switching times for the sequence of $34$ exchange interactions of Figure 3, given in units of $2 \hbar/ J$.}
\label{tab:4qbcnot}
\end{table}

Note that those exchanges that are on a vertical line in Figure
3 involve disjoint sets of qubits and can be applied
{\em in parallel}, where each cycle of gates lasts as long as the longest switching time in the set of parallel gates. If we count the number of parallel operations, we
obtain $19$ gate cycles.

\begin{figure}[ht]
\centering
\includegraphics{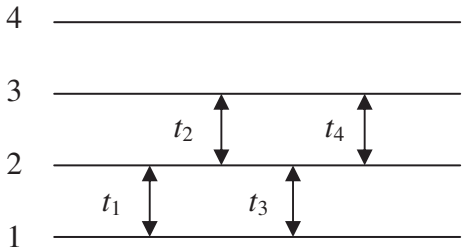}
\caption{Local unitaries can be generated using a sequence of $4$ exchange gates as shown.}
\label{fig:4qbUV4}
\end{figure}

In the  second stage we searched for the encoded local unitary gates $U_1,U_2,V_1,V_2$ (see Eq. (\ref{locequivalent})) that
transform the $34$-gate sequence into an exact $CNOT$ on the computational basis states, Eqs.~\ref{E:encoding1} and~\ref{E:encoding2}. It has been shown previously \cite{Bacon:99b,DiVincenzo:00a}
that each encoded local unitary can be obtained by a sequence of four
exchange gates. We employ here the nearest neighbor layout as shown in Figure
\ref{fig:4qbUV4}.
The constructions in \cite{Bacon:99b,DiVincenzo:00a} involve non-nearest neighbor interactions, with $E_{13}$ instead of $E_{23}$. It is easy to see, however,  that replacing $E_{13}$ with $E_{23}$ in the arguments of  \cite{Bacon:99b,DiVincenzo:00a} also leads to a sequence of four exchanges.

 These $16$ remaining gates ($4$ for each local unitary) can be
obtained either as the result of numerical optimization of a suitable
cost function, or from solving the system of non-linear equations
derived from the element-wise equivalency condition between the
objective matrix and the product of the four exchange matrices. We
employed the optimization approach using a Nelder-Mead simplex search
because of its high efficiency and generality.
 A similar approach has been used in \cite{Myrgren:03a}, where the cost function (a combination of a matrix distance between the actual gate and the desired gate and a non-leakage requirement) and the details of our numerical search calculations can be found. To reduce the probability of sampling only local minima, we sampled a large number (5 million) of randomly selected initial points in parameter space. We were able to determine the four local unitaries and their corresponding $4$-gate exchange sequences to a precision of $10^{-4}$ in the cost function, with the corresponding maximum  matrix element distance of the order of $10^{-5}$.

An alternative approach to finding gate
sequences for  local unitaries using the standard  mapping from $SU(2)$ to
$SO(3)$ and a quaternion representation of $SO(3)$ can be found in
Ref.~\cite{Myrgren:03a}.

 Table~\ref{tab:localunitaries} shows the gate times for each of the 4 exchange interactions required to implement the encoded local unitary operations.
\begin{table}[ht]
\begin{tabular}{|c|c|c|c|c|c|c|}\hline 
Exchange & Qubit  & Qubit &   U1  &   U2  &   V1  &   V2  \\
Time   &    1   &   2   & Times & Times & Times & Times \\
\hline \hline
$t_1$      &    1   &   2   & 2.218823 & 1.391831 & 4.865658 & 0.933012 \\
$t_2$      &    2   &   3   & 4.386508 & 1.977325 & 3.141319 & 2.025429 \\
$t_3$      &    1   &   2   & 3.442139 & 2.974488 & 1.493938 & 1.315318 \\
$t_4$      &    2   &   3   & 1.808165 & 2.105277 & 3.141314 & 0.042865 \\
\hline
\end{tabular}
\vskip.1cm
\caption{The exchange gates and corresponding gate times (in units of $2 \hbar/ J$) required to transform $W$ into the actual $CNOT$ gate.}
\label{tab:localunitaries}
\end{table} 
Note that the exchange interactions implementing $U_1$ and $U_2$ can
be applied in parallel, as can those for $V_1$ and $V_2$. Thus,
transforming the 34-gate sequence of Fig.
3 into the
exact $CNOT$ gate on the computational basis requires $16$ additional
nearest-neighbor 
interactions, that can be realized as $8$ additional parallel gate cycles.

The total number of nearest neighbor interactions for the exact $CNOT$
amounts to $27$ if applied in parallel and $50$ if applied serially.

\section{Gate sequences for the three-qubit encoding}

A $19$ exchange gate sequence for a gate locally equivalent to $CNOT$ for the three qubit encoding has been given in \cite{DiVincenzo:00a}. The layout in \cite{DiVincenzo:00a} obeyed certain symmetry constraints. We have recalculated this sequence to a higher accuracy with and without these symmetry constraints and have also computed the exchange-only implementation of the local unitaries $U_i,V_i$, $i=1,2$, needed to transform the $19$-gate sequence to the exact $CNOT$ in the computational basis. The resulting sequences for exact $CNOT$ in the three-qubit encoding are given in Appendix \ref{App:3}.

\section{Conclusion}

We have presented an exact construction of a  discrete universal logical
gate-set using only the exchange operation with a four-qubit encoding.  These results are
readily applicable to physical implementation schemes in which
exchange interactions are favored.  These include the classical solid-state nuclear-spin qubit
model proposed by Kane \cite{Kane:98a,Kane:00a} and the electron-spin qubit proposal of Loss and DiVincenzo \cite{Loss:98a,Burkard:99b}. For a four-fold
increase in the number of system qubits and a twenty-nine-fold
increase in the number of computational cycles for the two-qubit
operation, we are able to simplify the implementation of spin coupled
solid state systems by entirely eliminating the need for
single-spin \textit{A}-gates. In contrast to the rapid and
relatively easily-tunable two-spin \textit{J}-gates, the
\textit{A}-gates demand considerably greater device complexity and
$g$-factor engineering on solid-state heterostructures 
\cite{Vrijen:00a,DiVincenzo:99a}.

Thus far, explicit constructions of universal logical gate-sets for
exchange-only quantum circuits have been given on three-qubit
encodings \cite{DiVincenzo:00a} and for the four-qubit scheme
presented here. It should be noted that in principle the overhead in
spatial resources can be made arbitrarily small: asymptotically the
rate of encoding into subsystems converges to unity
\cite{Kempe:01a}. However we have to carefully evaluate the trade-offs
in space and time for each encoding. So far no constructive analytical
methods to lower bound the number of nearest neighbor interactions for
encoded gates exist. Using numerical methods yields an increase from
$19$ to $34$ gates for a gate equivalent to the encoded $CNOT$ going
from a three-qubit \cite{DiVincenzo:00a} to the present four-qubit
encoding. This seems to indicate that the rate of growth of the number
of nearest neighbor gates needed is rather large and that it is
probably wise to stick to small encodings if the error correcting
properties are not also to be incorporated.  However, we note that
these are all numerical solutions and are not guaranteed to be
optimal.  It would therefore be useful to obtain analytic bounds on
the minimum number of exchange gates required for encoded operations.

Another open problem is the application of encoded universality
to other interactions encountered in nature and in the laboratory, to
facilitate the search towards optimal physical schemes for implementation of universal quantum computation. We believe that the scheme presented here
provides a step in this direction and alleviates the task of the
quantum engineer working towards spin-coupled solid state quantum computation.

\section*{Acknowledgements}
The effort of the authors is sponsored by the Defense Advanced
Research Projects Agency (DARPA) and Air Force Laboratory, Air Force
Materiel Command, USAF, under agreement number F30602-01-2-0524, and
by DARPA and the Office of Naval Research under grant number
FDN-00014-01-1-0826.

\bibliographystyle{prsty}

\appendix
\section{The Makhlin Invariants}\label{App:1}

We give a brief description of how to calculate the {\em Makhlin invariants} \cite{Makhlin:00a} for an encoded two-qubit operation $W$. These invariants characterize a two-qubit operation up to equivalence by local unitaries (see Eq. (\ref{locequivalent})).  

In a first step  project the physical operator $W$ onto the logical
subspace:
\begin{equation}
M = P^{\dag} U_{cnot}^{exchange} P.,\label{E:WtoM}
\end{equation}
P is a 256-by-4 matrix whose column vectors are the basis states
\{$|0_{L}0_{L}\rangle$, $|0_{L}1_{L}\rangle$,
$|1_{L}0_{L}\rangle$, $|1_{L}1_{L}\rangle$\}, and $M$ is a matrix in
$SU(4)$. We next transform $M$ into the ``Bell-basis'' as
$M_{B} =Q^\dagger M Q$, where
\begin{equation}
Q = \frac{1}{\sqrt{2}}\left( \begin{array}{cccc} 1 & 0 & 0 & i\\
0 & i & 1 & 0\\0 & i & -1 & 0\\ 1 & 0 & 0 & -i\\
\end{array} \right).   \label{E:BellTrans}
\end{equation}
Finally, we define $m = M_{B}^{T} M_{B}$, to obtain the invariants $M_{1} = tr^{2}(m)/16 det M$ and $M_{2}= (tr^{2}(m) -
tr (m^{2}))/4 det M$.
For gates that are locally equivalent to the $CNOT$, $M_{1} = 0$ and
$M_{2} = 1$.

\section{Numerical search for a gate locally equivalent to $CNOT$}\label{App:2}

To obtain a gate $U_{cnot}^{exchange}$ which is locally equivalent to the encoded
$CNOT$, we applied a combination of genetic algorithms and Nelder-Mead
simplex searches. At the beginning of every search, we fixed a
sequence of qubit pairs to be coupled with an exchange interaction,
and optimized the fitness function $\mathcal{F}$ with respect to time
parameters only.  No restrictions on symmetry were imposed, unlike \cite{DiVincenzo:00a} (see Appendix \ref{App:3}). We started with a small number of couplings and
incremented the number of exchange interactions after each unsuccesful
attempt to find a gate equivalent to the $CNOT$. The final layout of the
exchanges is indicated in Figure 3. We found that
space generated by $\mathcal{F}$ was sufficiently complex such that
allowing the sequence of qubit-pairs to vary during the optimization
only introduced unnecessary complications into the search.

Even with the incorporation of the leakage parameter $\Lambda$ into
fitness function $\mathcal{F} = f + \Lambda$, the large space of
parameters is still marked with many local minima. Therefore, the
first stage of our search was a \textit{genetic algorithm}, whose
heuristic is well-equipped to score large spaces aggressively in order
to identify basins in which a global minimum may occur. Whereas
algorithms based on the hill-descent heuristic often trap themselves
into basins of local minima, genetic algorithms are able to traverse 
rapidly through regions of the space between the basins, enabling them
to descend from one basin to another. The pseudocode for the genetic
algorithm is as follows:

\textbf{\textsf{Step 1: Initialization.}}  Let the initial
population consisting of 60 \emph{candidates} be defined as the
set $\textbf{P}_{t}$. Each member of $\textbf{P}_{t}$ is a 34-dimensional 
real-valued vector whose elements lie in the interval [0, 2$\pi$].
Each vector represents the \textit{genome} of a candidate, and the
$j^{th}$ element in each vector ( a {\em gene}) is the time parameter for the
Hamiltonian in the $j^{th}$ exchange in a fixed sequence of
qubit-pair exchanges.

\textbf{\textsf{Step 2: Fitness Evaluation and Selection.}}
Generate the Makhlin invariants $M_{1}, M_{2}$ and leakage
parameter $\Lambda$ for each candidate and rank the candidates
according to their fitness scores $\mathcal{F}$. Sort the top $20$ performing
candidates into the \textit{parental pool}.

\textbf{\textsf{Step 3: Crossover.}}  Randomly pair the 20 members
of the parental pool. Each parental pair generates two offspring.
The first offspring is a random, pairwise convex combination of
the genomes of the parental pair. Let $\alpha$ and $\beta$ be
random real variables in the interval $[0,1]$. For parents \emph{u}
and \emph{v}, we have:
\begin{eqnarray} \label{E:ConvexCO}
PARENT_{u} &=& (\gamma_{u,1},\dots, \gamma_{u,34}) \nonumber\\
PARENT_{v} &=& (\gamma_{v,1},\dots, \gamma_{v,34}) \nonumber\\
OFFSPRING1_{u,v} &=& (\alpha\gamma_{u,1} + (1-\alpha)\gamma_{v,1},
\dots, \alpha\gamma_{u,34} + (1-\alpha)\gamma_{v,34})
\end{eqnarray}
The second offspring is a random geometric average of the genomes
of the parental pair.
\begin{eqnarray} \label{E:GeoCO}
PARENT_{u} &=& (\gamma_{u,1},\dots, \gamma_{u,34}) \nonumber\\
PARENT_{v} &=& (\gamma_{v,1},\dots, \gamma_{v,34}) \nonumber\\
OFFSPRING2_{u,v} &=& ({\gamma_{u,1}})^{\beta} ({\gamma_{v,1}})^{1 -
\beta}, \dots, ({\gamma_{u,34}})^{\beta} ({\gamma_{v,34}})^{1 -
\beta}
\end{eqnarray}
Intuitively, each candidate in the population represents a point
on a simplex within the search space.  By taking convex
combinations and geometric averages between the points, we search
the face planes of the simplex.

\textbf{\textsf{Step 4: Insertion.}}  We now construct the
population of the next generation $\textbf{P}_{t+1}$.  The new
population consists of:
\begin{enumerate}
\item The top (20 + \emph{M}) candidates from $\textbf{P}_{t}$,
where \emph{M} is a randomly generated integer between $0$ and $20$
\item The 20 offspring generated from the crossover step \item ($20$
- \emph{M}) new, randomly-generated candidates
\end{enumerate}
The purpose in inserting new candidates during each generation is
to enable the search to extract itself from local minima. If the
search simplex has converged to a local minimum, the new
candidates will serve as vertex points that can pull the search
into more promising regions within the space.

\textbf{\textsf{Step 5: Mutation.}}  We now subject the population to
a random mutation process, where each gene (component of
$\textbf{P}_{t}$) in each genome is perturbed to a new value within
$[0, 2 \pi]$ with probability $.03$.  It is necessary to introduce these
mutations, corresponding to small steps in the search simplex, because
the cross-over operations tend to pass over global minima too
rapidly. However, even small perturbations in the genome cause
increasingly violent movements in $\mathcal{F}$ as global minima are
approached. So to balance these considerations, the top ten performers
in each generation are exempted from mutations to stabilize the
performance of the algorithm.

\textbf{\textsf{Step 5': Exit Condition Check.}}  Check if the
top-ranked candidate satisfies the condition $\mathcal{F} < \epsilon$ for
a sufficiently small $\epsilon$. If not, return to Step 2.

We ran four simultaneous genetic algorithms with four distinct,
randomly-generated populations of size $60$, and coordinated the
search by inserting a clone of the top candidate from the population
with the best top-performer into the other three populations.  After
2394 generations, we obtained a candidate with error magnitudes of
$O(10^{-2})$ with respect to the Makhlin invariants and $O(10^{-1})$
with respect to the leakage parameter.  At this point, the pace of
progress in the genetic algorithm slowed down dramatically, so we used
the top performer as the starting point for the second stage of our
algorithm, a Nelder-Mead simplex direct search. At this point, the
simplex search was considerably more robust, because the simplex
heuristic enables the simplex to flex and squeeze itself through
narrow valleys of the space more sensitively. After 5296 iterations of
the simplex search, we obtained a candidate with error magnitudes of
$O(10^{-6})$ with respect to the Makhlin invariants and $O(10^{-2})$
with respect to the leakage parameter. Since this was the first time
we had advanced to such a low point in the space, we ran the genetic
algorithm again to see if any further improvement could be obtained in
this manner, and to acquire some further intuition about the structure
of the space.  After 471 generations, only a trivial improvement was
obtained, so we returned to the simplex method once more. After 22081
iterations of the simplex search, we obtained error magnitudes of
$O(10^{-10})$ with respect to the Makhlin invariants and $O(10^{-8})$
with respect to the leakage parameter.

\section{Gate sequences for $CNOT$ with the three-qubit encoding}\label{App:3}

In Ref.~\cite{DiVincenzo:00a}, a numerical search was utilized to 
generate
a $CNOT$ gate from a sequence of nearest-neighbor exchange interactions
on a system of two computational qubits, each encoded by three
physical spin-1/2 qubits. The three-qubit encoding is
\begin{align}\label{E:NatureEncoding}
|0_{L} \rangle &= \frac{1}{2}|S \rangle \otimes | 1 \rangle \nonumber\\
|1_{L} \rangle &= \sqrt{\frac{2}{3}}|T_{+} \rangle \otimes |0
\rangle - \frac{1}{\sqrt{3}}|T_{0} \rangle \otimes | 1 \rangle,
\end{align}
where $|S \rangle = \frac{1}{\sqrt{2}}\left( |10 \rangle - | 01
\rangle \right)$, $|T_{+} \rangle = |11 \rangle$ and $|T_{0}
\rangle = \frac{1}{\sqrt{2}}\left( |10 \rangle + | 01 \rangle
\right)$. The numerical search was made by minimization of a similar fitness
function to that employed here, {\it i.e.}, including both matrix distance 
from $CNOT$ and leakage penalty functions.
 A candidate sequence of $19$ exchanges  equivalent to $CNOT$
up to local transformations on the encoded basis ($U_{cnot}^{exchange}$) was obtained for a layout
containing two simplifying symmetries in the exchange gate times.  These symmetries are illustrated in Figure 5, with the gate switching times given in Table 3. Here we give also the exchange-only implementation of the local gates $U_i,V_i,$ $i=1,2$ (cf. Eq. (\ref{locequivalent})), needed to convert the $19$-gate sequence into the exact $CNOT$ in the computational basis.  These were obtained using the same procedures as described in Section~\ref{sec:CNOT} and Ref.~\cite{Myrgren:03a}. 

\begin{figure}[ht]\label{fig:appendix1}
\begin{picture}(0,220)
\includegraphics{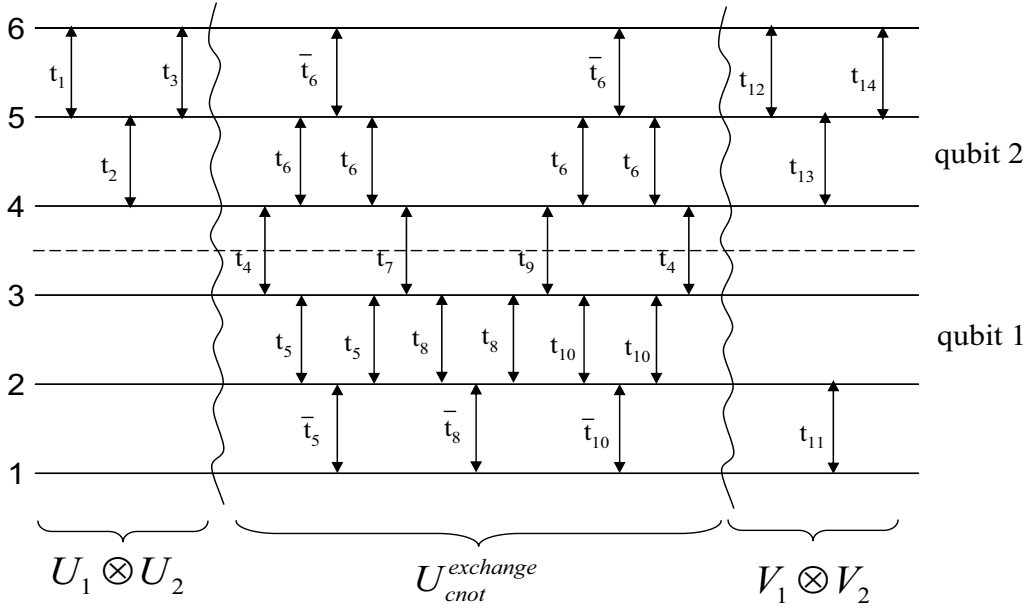}
\end{picture}
\caption{ Layout for the exact $CNOT$. The 19-gate sequence for a gate $U_{cnot}^{exchange}$ locally equivalent to $CNOT$ ($t_4-t_{10}, \bar{t}_5-\bar{t}_{10}$) is taken from Ref. \cite{DiVincenzo:00a} (time indices are shifted by three). Gate switching times are given in Table 3.}
\end{figure}

\begin{table}[t] 
\label{Table:another}
\begin{tabular}{|c|c|c|c|c|c|c|c|}\hline 
Exchange & Qubit  & Qubit &          & Exchange & Qubit & Qubit & \\
Time   &    1   &   2   & Times     & Time   &   1   &   2   & Times \\
\hline \hline
$t_1$    &    5   &   6   & 0.863060  & $t_8$  &    2   &   3   & 1.302881 \\
$t_2$    &    4   &   5   & 0.303496  & $t_9$  &    3   &   4   & 0.463869 \\
$t_3$    &    5   &   6   & 0.863060  & $t_{10}$  &    2   &   3   & 2.554511 \\

$t_4$    &    3   &   4   & 1.290877  & $t_6$  &    4   &   5   & 0.871873 \\
$t_5$    &    2   &   3   & 0.650655  & $\bar{t_{10}}$  &    1   &   2   & 1.249644 \\
$t_6$    &    4   &   5   & 0.871873  & $\bar{t_6}$  &    5   &   6   & -1.034121 \\
$\bar{t_5}$    &    1   &   2   & -1.207108 & $t_{10}$  &    2   &   3   & 2.554511 \\
$\bar{t_6}$    &    5   &   6   & -1.034121 & $t_6$  &    4   &   5   & 0.871873 \\
$t_5$    &    2   &   3   & 0.650655  & $t_4$  &    3   &   4   & 1.290877 \\
$t_6$    &    4   &   5   & 0.871873  & $t_{11}$  &    1   &   2   & 0.612497 \\
$t_7$ &    3   &   4   & 2.012205  & $t_{12}$  &    5   &   6   & 2.826113 \\
$t_8$ &    2   &   3   & 1.302881  & $t_{13}$  &    4   &   5   & 2.838096 \\
$\bar{t_8}$ &    1   &   2   & -0.502098 & $t_{14}$  &    5   &   6   & 2.278532 \\
\hline
\end{tabular}
\vskip.1cm
\caption{Gate switching times for the $26$ gate sequence of Figure 5, given in units of $2 \hbar/J$. This corresponds to $\pi$ times the time units employed in Ref. \cite{DiVincenzo:00a}.}
\end{table}

The first symmetry is that certain gate times are repeated in a spatially
symmetric configuration, {\it e.g.}, $t_4$ occurs at the beginning and end of the sequence
between physical qubits $3$ and $4$ in each case. We denote this
the \textit{repetition symmetry}.  A second symmetry apparent in Figure 5 is
that certain sequential pairs of interaction times are
related by analytic functions. In particular, for $k=5,6,8,10$, the functions
\begin{equation}\label{E:TbarRelation}
c_{k} = \textup{tan}(t_{k})\textup{tan}(\bar{t}_{k}) + 2
\end{equation}
are exactly equal to zero.  We denote this symmetry a \textit{correlation
symmetry}. 
With these symmetry restrictions the optimal solution for a $19$ exchange gate sequence 
obtained in Ref~\cite{DiVincenzo:00a}
yields a Makhlin invariant fitness value of 
 $f$  on the order of $10^{-10}$ and a
 leakage parameter $\Lambda$ on the order of $10^{-8}$.   The overall
precision of the sequence, given by the maximum matrix element distance to the $CNOT$ in
the computational basis is of order $10^{-6}$.

We have 
performed a new set of numerical searches for this same $19$-exchange gate layout in the three-qubit encoding,
using the techniques described in this paper and without imposing any
symmetry constraints on the exchange times.  We find that not only can higher
quality numerical solutions be
obtained, but also that these are very significantly improved when the simplifying symmetry constraints are lifted.  This
results in  a larger set of independent variables ($19$ instead of the $7$ given in Ref.~\cite{DiVincenzo:00a})
but with the advantage of a considerably smaller value of the cost function. 

 We performed a search with
a Nelder-Mead simplex algorithm whose optimization criterion was
the minimization of the Makhlin fitness function $f$ and the
leakage parameter $\Lambda$, starting with a random initial set of exchange gate times. 
Figure 6 and Table 4 shows the corresponding exchange sequence, together with the exchange-only implementation of the local gates $U_i, V_i$, $i=1,2$ needed to
convert the $19$-gate sequence into the exact $CNOT$ in the computational basis.  This search yielded zero values of Makhlin fitness function
$f$ and leakage parameter $\Lambda$ to within machine precision ($10^{-16}$), and a maximum matrix element distance from the exact $CNOT$ of $10^{-9}$. This provides a significant improvement over the corresponding overall precision of $10^{-6}$ for the original $19$-gate sequence of Ref.~\cite{DiVincenzo:00a}.
\begin{figure}[ht]\label{fig:appendix2}
\begin{picture}(0,240)
\includegraphics{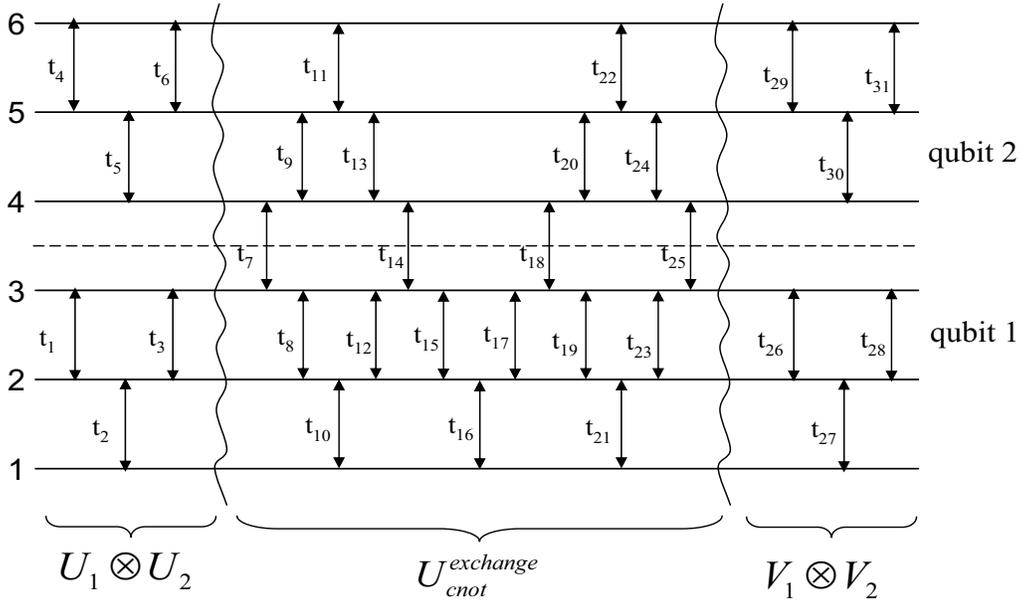}
\end{picture}
\caption{Layout for the exact $CNOT$ obtained by unrestricted optimization of times $t_7-t_{25}$ for $U_{cnot}^{exchange}$ and of $t_1-t_6$ and $t_{26}-t_{31}$ for the local unitaries $U_i,V_i$, $i=1,2$. Gate switching times are given in Table 4.}
\end{figure}

\begin{table}[t]
\label{Table:2}
\begin{tabular}{|c|c|c|c|c|c|c|c|}\hline 
Exchange & Qubit  & Qubit &           & Exchange & Qubit & Qubit & \\
Time   &    1   &   2   & Times      & Time   &    1  &   2   & Time \\
\hline \hline
$t_1$    &    2   &   3   & 3.141592  & $t_{17}$  &    2  &   3   & 4.444461 \\ 
$t_2$    &    1   &   2   & 0.989737  & $t_{18}$  &    3  &   4   & 0.463873 \\
$t_3$    &    2   &   3   & 3.141593  & $t_{19}$  &    2  &   3   & 1.249608 \\
$t_4$    &    5   &   6   & 2.477807  & $t_{20}$  &    4  &   5   & 5.249065 \\
$t_5$    &    4   &   5   & 0.303496  & $t_{21}$  &    1  &   2   & 2.554454 \\
$t_6$    &    5   &   6   & 0.863060  & $t_{22}$  &    5  &   6   & 4.013466 \\
$t_7$    &    3   &   4   & 4.432470  & $t_{23}$  &    2  &   3   & 4.391200 \\
$t_8$    &    2   &   3   & 3.792238  & $t_{24}$  &    4  &   5   & 2.107472 \\
$t_9$    &    4   &   5   & 2.107472  & $t_{25}$  &    3  &   4   & 1.290877 \\
$t_{10}$ &    1   &   2   & 5.076069  & $t_{26}$  &    2  &   3   & 3.141592 \\
$t_{11}$ &    5   &   6   & 0.871873  & $t_{27}$  &    1  &   2   & 0.927636 \\
$t_{12}$ &    2   &   3   & 3.792237  & $t_{28}$  &    2  &   3   & 3.141592 \\
$t_{13}$ &    4   &   5   & 5.249065  & $t_{29}$  &    5  &   6   & 0.863060 \\
$t_{14}$ &    3   &   4   & 5.153789  & $t_{30}$  &    4  &   5   & 0.303496 \\
$t_{15}$ &    2   &   3   & 1.302870  & $t_{31}$  &    5  &   6   & 0.466283 \\
$t_{16}$ &    1   &   2   & 5.781068  &           &        &       & \\ 
\hline
\end{tabular}
\vskip.1cm
\caption{Gate switching times for the $31$ gate sequence of Figure 6, given in units of $2 \hbar/ J$.} 
\end{table}
In the solution described in \cite{DiVincenzo:00a}, the
correlation symmetries were satisfied to machine precision, while
the repetition symmetries satisfied exactly. 
We note that these correlation and repetition symmetries are not essential to the
task of implementing the encoded $CNOT$ operation. From an
optimization perspective, they might even be interpreted as a
hindrance that constrains the trajectory of numerical searches to
lie in sub-optimal subspaces of the control parameter space.  It
is not clear whether the symmetries in the solution obtained in
Ref.~\cite{DiVincenzo:00a} suggest the existence of analytical
solutions to the optimization problem of the
cost function, and if so whether these correspond to local or global minima.
The above example shows that, without any symmetry restrictions and allowing the number of independent time
parameters to increase, improvement to optimization of the cost function to within machine prevision can be obtained.

\end{document}